\documentclass[8.5pt,twoside,twocolumn]{article}
\usepackage[super,sort&compress,comma]{natbib} 
\usepackage{mhchem}
\usepackage{times,mathptmx}
\usepackage{sectsty}
\usepackage{balance} 

\usepackage{graphicx} %eps figures can be used instead
\usepackage{lastpage}
\usepackage[format=plain,justification=raggedright,singlelinecheck=false,font=small,labelfont=bf,labelsep=space]{caption} 
\usepackage{fancyhdr}
\begin{document}

\thispagestyle{plain}
\fancypagestyle{plain}{
%\fancyhead[L]{\includegraphics[height=8pt]{headers/LH}}
%\fancyhead[C]{\hspace{-1cm}\includegraphics[height=20pt]{headers/CH}}
%\fancyhead[R]{\includegraphics[height=10pt]{headers/RH}\vspace{-0.2cm}}
\renewcommand{\headrulewidth}{1pt}}
\renewcommand{\thefootnote}{\fnsymbol{footnote}}
\renewcommand\footnoterule{\vspace*{1pt}% 
\hrule width 3.4in height 0.4pt \vspace*{5pt}} 
\setcounter{secnumdepth}{5}

\makeatletter 
\def\subsubsection{\@startsection{subsubsection}{3}{10pt}{-1.25ex plus -1ex minus -.1ex}{0ex plus 0ex}{\normalsize\bf}} 
\def\paragraph{\@startsection{paragraph}{4}{10pt}{-1.25ex plus -1ex minus -.1ex}{0ex plus 0ex}{\normalsize\textit}} 
\renewcommand\@biblabel[1]{#1}            
\renewcommand\@makefntext[1]% 
{\noindent\makebox[0pt][r]{\@thefnmark\,}#1}
\makeatother 
\renewcommand{\figurename}{\small{Fig.}~}
\sectionfont{\large}
\subsectionfont{\normalsize} 

\fancyfoot{}
%\fancyfoot[LO,RE]{\vspace{-7pt}\includegraphics[height=9pt]{headers/LF}}
%\fancyfoot[CO]{\vspace{-7.2pt}\hspace{12.2cm}\includegraphics{headers/RF}}
%\fancyfoot[CE]{\vspace{-7.5pt}\hspace{-13.5cm}\includegraphics{headers/RF}}
\fancyfoot[RO]{\footnotesize{\sffamily{1--\pageref{LastPage} ~\textbar  \hspace{2pt}\thepage}}}
\fancyfoot[LE]{\footnotesize{\sffamily{\thepage~\textbar\hspace{3.45cm} 1--\pageref{LastPage}}}}
\fancyhead{}
\renewcommand{\headrulewidth}{1pt} 
\renewcommand{\footrulewidth}{1pt}
\setlength{\arrayrulewidth}{1pt}
\setlength{\columnsep}{6.5mm}
\setlength\bibsep{1pt}

\twocolumn[
  \begin{@twocolumnfalse}
\noindent\LARGE{\textbf{Elasticity-Dependent Self-assembly of Micro-Templated Chromonic Liquid Crystal Films$^\dag$}}
\vspace{0.6cm}

\noindent\large{\textbf{Matthew A. Lohr,$^{\ast}$\textit{$^{a}$} Marcello Cavallaro Jr.,\textit{$^{b}$} Daniel A. Beller,\textit{$^{a}$} Kathleen J. Stebe,\textit{$^{b}$} Randall D. Kamien,\textit{$^{a}$} Peter J. Collings,\textit{$^{a, c}$}and
Arjun G. Yodh\textit{$^{a}$}}}\vspace{0.5cm}
%Please note that \ast indicates the corresponding author(s) but no footnote text is required. 

\vspace{0.6cm}
%Please do not change this text.

\noindent \normalsize{We explore micropatterned director structures of aqueous lyotropic chromonic liquid crystal (LCLC) films created on square-lattice cylindrical-micropost substrates. The structures are manipulated by modulating the LCLC mesophases and their elastic properties via concentration through drying. Nematic LCLC films exhibit preferred bistable alignment along the diagonals of the micropost lattice. Columnar LCLC films, dried from nematics, form two distinct director and defect configurations: a diagonally aligned director pattern with local squares of defects, and an off-diagonal configuration with zig-zag defects. The formation of these states appears to be tied to the relative splay and bend free energy costs of the initial nematic films. The observed nematic and columnar configurations are understood numerically using a Landau-de Gennes free energy model. Among other attributes, the work provide first examples of quasi-2D micropatterning of LC films in the columnar phase and lyotropic LC films in general, and it demonstrates alignment and configuration switching of typically difficult-to-align LCLC films via bulk elastic properties.}
\vspace{0.5cm}
 \end{@twocolumnfalse}
  ]

%\textit{e.g.} [Surname \textit{et al., Journal Title}, 2000, \textbf{35}, 3523].

\section{Introduction}
%Footnotes
\footnotetext{\dag~Electronic Supplementary Information (ESI) available.}

%Please use \dag to cite the ESI in the main text of the article.
%If you article does not have ESI please remove the the \dag symbol from the title and the above footnotetext.

\footnotetext{\textit{$^{a}$~Department of Physics and Astronomy, University of Pennsylvania, Philadelphia, Pennsylvania, 19104, USA.}}
\footnotetext{\textit{$^{b}$~Department of Chemical and Biomolecular Engineering, University of Pennsylvania, Philadelphia, Pennsylvania, 19104, USA. }}
\footnotetext{\textit{$^{c}$~Department of Physics and Astronomy, Swarthmore College, Swarthmore, Pennsylvania, 19081, USA. }}

%additional addresses can be cited as above using the lower-case letters, c, d, e... If all authors are from the same address, no letter is required

%\footnotetext{\ddag~Additional footnotes to the title and authors can be included \emph{e.g.}\ `Present address:' or `These authors contributed equally to this work' as above using the symbols: \ddag, \textsection, and \P. Please place the appropriate symbol next to the author's name and include a \texttt{\textbackslash footnotetext} entry in the the correct place in the list.}

The engineering of novel metamaterials and biomimectic structures demands increasingly creative methods for shaping materials at the microscale. In recent years, self-assembly has been explored as a route to the creation of complex structure in various soft materials.  Promising methods of self-assembly include programmed assembly of colloidal particles with designed interactions \cite{Feng2013, McGinley2013} and controlled buckling in thin films due to competing stresses .\cite{Bowden2008, Yin2012, Kim2012}  In the latter case, the interplay of bulk elastic properties guided by structured templates can give rise to varied and complex microstructures. \cite{Bowden2008}

In a similar vein, self-assembly via templating of elastically anisotropic media has recently been demonstrated in liquid crystal films confined by micropillar arrays.  Some such experiments employed microstructured templates to create arrays of defects in liquid crystal films.\cite{Honglawan2011, Honglawan2013, Cavallaro2013}  Furthermore, these structures have been shown to be useful for the nucleation of novel bulk (3D) LC phases\cite{Honglawan2011, Honglawan2013} and colloidal configurations at surfaces.\cite{Cavallaro2013}  Related work has induced bistable local director alignment in liquid crystals, yielding applications for low-energy display technologies.\cite{Kitson2002, Kitson2004, Kitson2008}  However, to date such methods have only been applied to \textit{thermotropic} nematic and smectic liquid crystals.

Lyotropic chromonic liquid crystals (LCLCs) are an important and relatively unexplored class of anisotropic fluids conducive to microscale self-assembly and compatibile with aqueous media.  LCLCs are typically composed of molecular aggregates of plank-like polyaromatic compounds with ionic side-groups.  As a result,  their mesophases differ significantly from thermotropic and amphiphilic lyotropic liquid crystals.    Patterned films of LCLCs have a wide variety of emerging applications distinct from other types of liquid crystals, including inexpensive polarizing films,\cite{Dreyer1948, Lydon2010, Lydon2011} holographic displays,\cite{TamChang2006, Matsunaga2002} organic electronics and solar cells,\cite{Nazarenko2010APL, ONeill2011} biosensors,\cite{Shiyanovskii2005, Woolverton2005}, aqueous colloidal, nanotube and bacterial assembly,\cite{Evans2011, OuldMoussa2013, Smalyukh2008, Mushenheim2014, Zhou2014} and precursors to structured graphene-based materials.\cite{Guo2011, Mukhopadhyay2013} LCLCs also offer useful attributes for fundamental investigation of the effects of elasticity on self-assembly behavior, since their elastic properties can be tuned via control of mesogen concentration,\cite{Zhou2012} depletants and ions.\cite{Park2011} Indeed, studies have demonstrated that LCLCs in micro-scale confinement form unique, elastic-property-dependent configurations.\cite{Tortora2011, Jeong2014} 

In this contribution we manipulate the patterning of LCLC films within micropost arrays by modulating the LCLC mesophases and elastic properties via concentration changes during a drying process.  We discover a preferred alignment in nematic films governed by elastic interactions with post-placement geometry, and we observe multiple fascinating stable patterned states of the columnar phase.  This templating scheme has not been applied previously to lyotropic LC systems, and this contribution provides a first example of quasi-2D micropatterning of columnar LC films.  We develop a numerical model to understand observed differences in patterning, and this model suggests that such effects are due to the inherent anisotropic elastic properties of films that evolve during a slow ``concentrating'' process.  Thus the multiple configurations that form are based on changes in bulk elastic properties, rather than changes in applied fields \cite{Kitson2002, Kitson2004} or changes in confinement geometry \cite{Kitson2008}.

\section{Experimental and Numerical Methods}

\subsection{Experimental Materials and Methods}

We create chromonic liquid crystal films in photolithographically printed arrays of cylindrical microposts (Figure 1).  Photolithographic techniques are used to make cylindrical microposts of negative tone epoxy-based photoresist, SU-8 2000 series (Microchem Inc.) on a glass substrate.  We employ cylindrical posts in order to separate the effects of micropost placement from micropost shape, since previous studies show that anisotropic or sharp-featured post shapes can dominate alignment and organization of liquid crystals.\cite{Thurston1980, Kitson2002} Micropost cylinders with height $h$ =5 $\mu$m, and diameter $d$ = 7$\mu$m, are arranged in a square lattice with center-to-center pitch of $a$ = 14$\mu$m.

\begin{figure}
\includegraphics[width=0.95\linewidth]{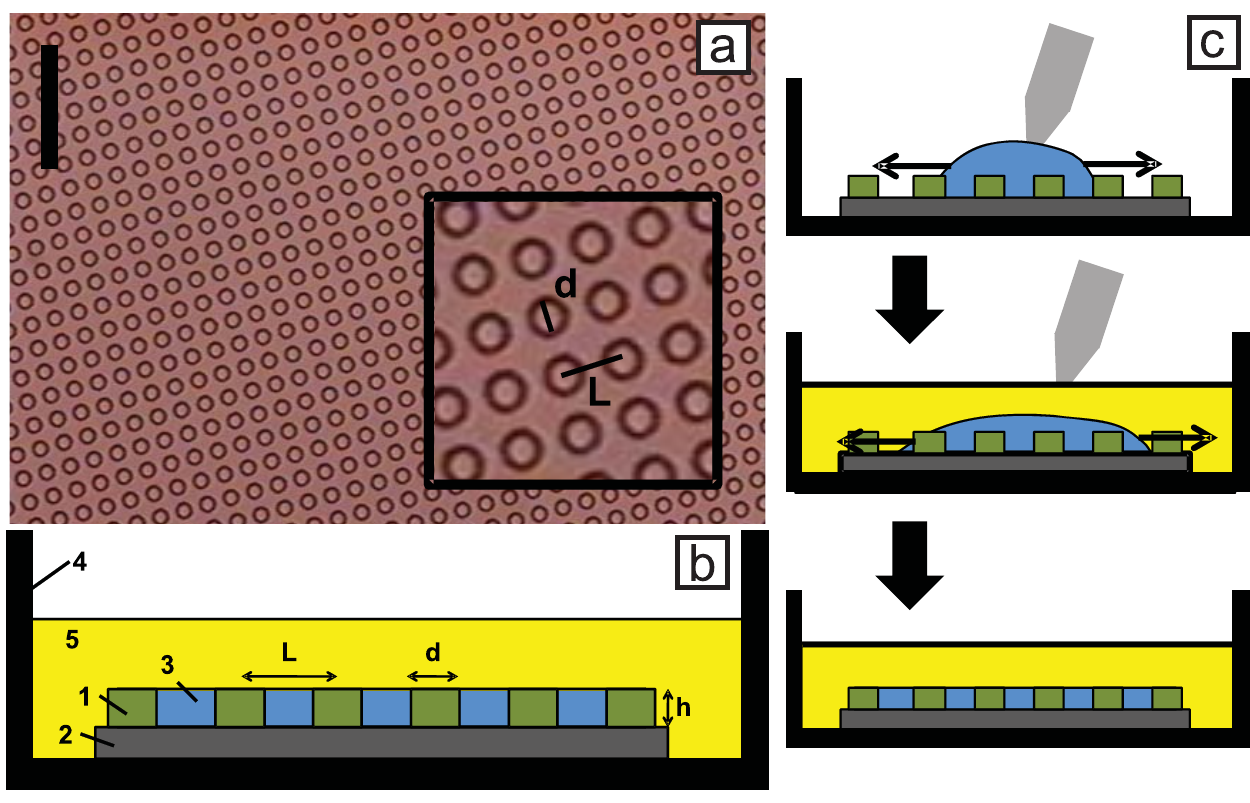}
\caption{Configuration of micropost arrays loaded with liquid crystals.  a) Bright-field microscopy image of an empty SU-8 micropillar array with post center spacing $L$ = 14$\mu$m, diameter $d$ = $L$/2 = 7$\mu$m, and height $h$ = 5$\mu$m.  Scalebar = 50 $\mu$m.  Inset: close-up of microposts, indicating $d$ and $L$.  b) Schematic of micropost array configuration after loading with LCLC film.  SU-8 microposts (1) sit atop a glass wafer (2).  After being loaded with a LCLC (3), the wafer is placed in a glass petri dish (4) and immersed in hexadecane(5).  c) Schematic of LCLC film loading procedure.  A droplet of heated LCLC suspension is placed in a droplet onto a cleaned and heated micropost array where it begins to spread due to capillary forces (top); immediately afterward, hexadecane is placed to completely cover the sample, and LCLC droplet continues to spread (middle); after several minutes, the LCLC film has spread to cover the entire micropost array, leaving a level film at the post height (bottom).}
\end{figure}

After the micropost arrays are thoroughly rinsed and plasma-cleaned, they are filled with a nematic aqueous LCLC suspension.  We use one of two LCLC-forming molecules suspended in Millipore filtered water at various concentrations: disodium cromoglycate (DSCG, Sigma Aldrich), and Sunset Yellow FCF (SSY, Sigma-Aldrich).  DSCG is sufficiently pure to be used as-is; SSY, by contrast, is purified by re-suspension in water, precipitation through the addition of ethanol, and centrifugation several times in order to reach a purity of $>$ 99 $\%$.  These particular LCLCs were selected because their elastic properties have been characterized thoroughly in previous studies. \cite{Nastishin2008, Zhou2012}

The procedure for filling the micropost array with LCLC is illustrated in Figure 1c.  Micropost-covered glass slides are placed in a petri dish and heated to 60 $^\circ$C. At the same time, the LCLC suspension (which is in the isotropic phase at this temperature) is also heated to 60 $^\circ$C.  A droplet of 2-3 $\mu$L of LCLC solution is placed on the substrate and then spreads rapidly due to capillary interactions with the microposts.  Several mL of hexadecane is then immediately placed over the LCLC and substrate in order to prevent rapid evaporation of the aqueous suspension.  The entire system is left at 60 $^\circ$C for several minutes, until the LCLC droplet has ceased spreading.  The petri dish is then placed on a microscope stage at 25 $^\circ$C and allowed to cool to ambient temperature, at which time the LCLC film re-enters the nematic mesophase.  The LCLC-hexadecane surface tension is large enough to insure that the aqueous film spreads evenly through the micropost array, eventually resting at the high-edge of the micropillars with an undeformed, flat interface.   Under these conditions, all surfaces in contact with the aqueous phase (glass substrate, pillar sides, and hexadecane interface) induce degenerate planar anchoring on the LCLC mesogens.

\subsection{Director Field Analysis of LCLC Films}

Typically, some variety of polarizing microscopy is used to gain information about director configurations in liquid crystals. Many studies employ qualitative analysis of birefringent samples between polarizing materials, but only recently have high-resolution director configurations been acquired from polarizing microscopy, though this approach typically requires advanced and costly supplementary imaging equipment. By contrast, in our work, we acquire high-resolution director configurations of our liquid crystal films using a relatively simple microscope setup that employs a combination of video microscopy, particle tracking and image analysis techniques.

Though this procedure is fully described in supplementary information, we briefly summarize the technique here.  Digital video is acquired of a liquid crystal film on a stage that can be manually rotated with respect to fixed cross polarizers.  We then track the micropost locations and motion using a combination of sub-pixel resolution tracking techniques.\cite{Crocker}  Importantly, from the tracked collective motion of the posts, we are able to subtract the global sample translation and rotation from the video microscopy data in order to derive ``effective'' images of a fixed micropost array between rotating cross-polarizers.  Then, using these rotated images, we fit the variation of pixel intensity in the LCLC with respect to cross-polarizer angle to the expected response of a birefringent material between cross-polarizers. This method allows us to extract the z-averaged planar-projection of the director field at each pixel in the microscope image.  Therefore, the spatial resolution of the director patterns is set by the pixel width, which, at the magnifications used in these experiments, is approximately 300 nm.

\subsection{Numerical Director Free Energy Minimizations}

To supplement and verify our experimental observations of liquid crystal configurations in micropost arrays, we perform numerical minimizations for a phenomenological Landau de-Gennes (LdG) free energy of a nematic director field under similar confinement conditions.\cite{Ravnik2009, Cavallaro2013} The free energy is minimized in a finite difference scheme on a regular cubic mesh, using a conjugate gradient minimization routine from the ALGLIB package.  Specifically, we simulate a box of 50 $\times$ 50 $\times$ 15 points, with a 24-point diameter cylindrical pillar in the center, tangential boundary conditions at the top, bottom and post surface, and periodic boundary conditions at the edges.  In the uniaxial limit, the LdG free energy is written in terms of the tensor $Q_{ij}=  \frac{3}{2} S(n_in_j - \frac{1}{3} \delta_{ij})$, where $n_i$ is the $i$th component of the nematic director, $\delta_{ij}$ is the Kronecker delta, and $S$ is the nematic degree of order. The free energy density (per unit area) is a sum of two components: a phase free energy density

\begin{equation} 
f_{\mathrm{phase}} = \int{dV\left(\frac{1}{2} A \mathrm{Tr}\left(\mathbf{Q}^2\right) + \frac{1}{3} B \mathrm{Tr} \left(\mathbf{Q}^3 \right) + \frac{1}{4} C \left(\mathrm{Tr}\left(\mathbf{Q}^2\right)\right)^2 \right)} 
\end{equation}
and a gradient free energy density, which, for a generic nematic LC with equal elastic constants reads

\begin{equation} 
f_{\mathrm{d}} = \frac{1}{2} L \int{dV\left( \nabla \mathbf{Q} \right)^2}.
\end{equation}
For lack of better information about elastic properties of LCLCs, we used values for A, B and C consistent with a commonly studied thermotropic nematic liquid crystal, 4-cyano-4'-pentylbiphenyl (5CB).

Though the above free energy suffices for qualitative modeling of generic nematic liquid crystals, applying these same numerical methods for understanding columnar configurations requires care. The distortion free energy $f_d$ shown in Equation 2 is a valid approximation for modeling a generic thermotropic nematic like 5CB. This can be re-stated as a Frank free energy,

\begin{equation}
f_{\mathrm{Frank}} = \frac{K_1}{2}(\nabla \cdot \textbf{n})^2 + \frac{K_2}{2}(\textbf{n} \cdot \nabla \times \textbf{n})^2 + \frac{K_3}{2}((\textbf{n} \cdot \nabla)\textbf{n})^2,
\end{equation}
where $K_1$, $K_2$, and $K_3$, the splay, twist, and bend elastic constants, respectively, are similar in magnitude.  However, as a nematic transitions into a columnar phase, there is a necessary coupling between the density and the director \cite{tarmey,kln,kt} that results in a lengthscale dependence of $K_1$, as well as a coupling between the crystalline order and the director, resulting in a lengthscale dependence of $K_2$.\cite{kn}  In both cases these elastic constants diverge at long wavelengths.  Moreover, even in the nematic phase, fluctuations near the nematic to columnar transition strongly renormalize the elastic constants.
\cite{Swift1982} Without implementing the full theory of the columnar phases we will consider the nematic theory with differing elastic constants and consider the limits where $K_1$ and $K_2$ grow in comparison to $K_3$.  The above form of the LdG free energy cannot model this system, as it assumes equal $K$'s.  However, if we use an expanded form of the LdG free energy\cite{Ravnik2009},

\begin{equation} 
%f_{\mathrm{d}} = \frac{1}{2} L H\left( \nabla \mathbf{Q} \right)^2.
f_{\mathrm{d}} = \frac{1}{2} \left( L_1 \frac{\partial Q_{ij}}{\partial x_k} \frac{\partial Q_{ij}}{\partial x_k} + L_2 \frac{\partial Q_{ij}}{\partial x_j} \frac{\partial Q_{ik}}{\partial x_k}+L_3 Q_{ij}\frac{\partial Q_{kl}}{\partial x_i} \frac{\partial Q_{kl}}{\partial x_j} \right),
\end{equation}

then, following previous numerical work\cite{Ravnik2009}, we find the following relationships between $L$ and $K$ terms:

%\begin{equation} 
\begin{eqnarray*}
K_1 = \frac{9}{4} S^2 \left( 2 L_1 + L_2 - S L_3 \right) \\
%\end{equation}
%\begin{equation} 
K_2 = \frac{9}{4} S^2 \left( 2 L_1 - S L_3  \right) \\
%\end{equation}
%\begin{equation} 
K_3 = \frac{9}{4} S^2 \left( 2 L_1 + L_2 + S L_3\right).
\end{eqnarray*}
%\end{equation}

Therefore, to model columnar configurations, we run the LdG free energy minimizations under the same conditions as for the nematic case, except we employ the expanded form of the LdG free energy of Equation 4, along with values of $L_i$ such that $K_3$ is decreased by at least an order of magnitude compared to $K_1$ and $K_2$.  Thus, even though the LdG free energy models a nematic LC, by using this limit of elastic constants, we can approximate an expected director field as a nematic approaches the columnar phase.  Further, instead of initializing the director configuration randomly (which leads to random defect nucleation and frustrated high energy states), we start with a bulk director field uniformly oriented in the plane of the post array.

\section{Results and Discussion}

\subsection{Nematic Configurations}

Initial observations of the nematic LCLC films under cross-polarizers indicate no obvious consistent director patterning on a scale larger than a single pillar spacing (though the director orientation appears to lie predominantly in-plane).  Considering that there may exist multiple metastable local director configurations for a nematic under such confinement, we sought to create a more consistently ordered film by introducing a weak directional bias into the film.  To achieve this goal, we form the micropillar arrays on glass surfaces that are first rubbed with a fine abrasive pad (3M Trizact Foam Disc P3000).  This rubbing creates nano-sized grooves in the bottom glass surface which, in turn, induce a weak, oriented planar alignment on the overlying chromonic liquid crystal \cite{Zhou2012, McGinn2013}.

\begin{figure}
\includegraphics[width=0.95\linewidth]{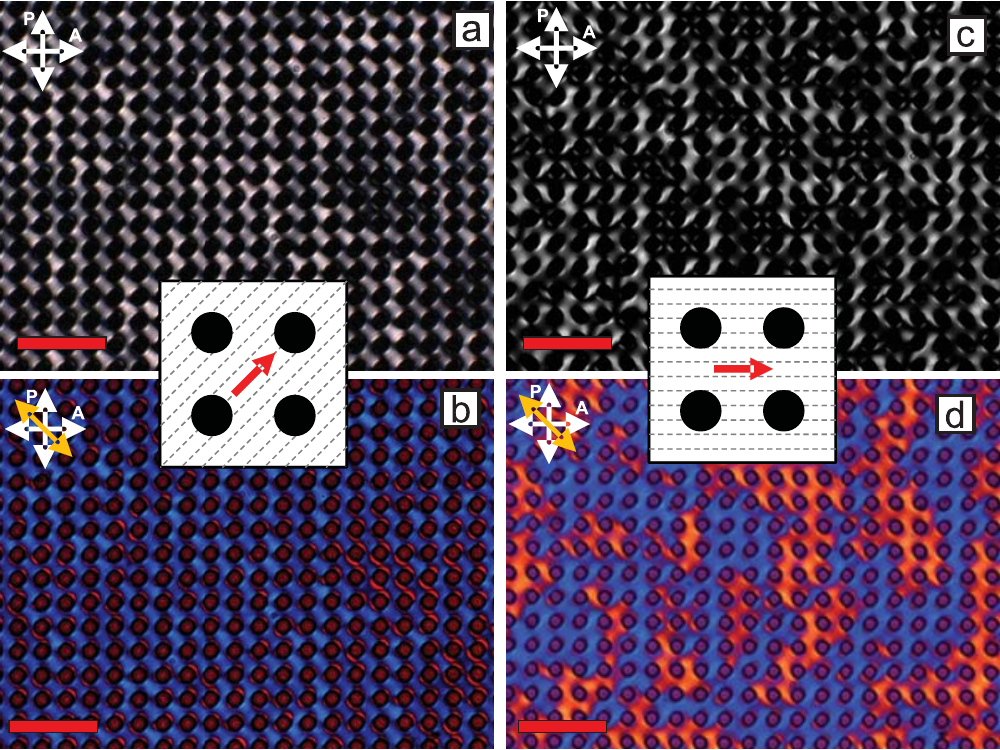}
\caption{Micropost arrays filled with nematic DSCG (14\% wt/wt) at room temperature.  Cross-polarizer (a \& c) and cross-polarizer with full wave retardation plate(b \& d) images shown with bottom surfaces rubbed along a lattice diagonal (a \& b) or a post lattice direction (c \& d), as illustrated in the insets.  Scale bar = 50 $\mu$m.  The slow axis of the retardation plate is given by the orange arrow in (b \& d).}
\end{figure}

As can be seen in Figure 2, the consistency of director patterning in the final film is dependent on the rubbing direction on the bottom surface.  For example, a surface rubbing along a diagonal of the post lattice results in a consistent nematic texture under cross-polarizers (2a, 2b).  The same degree of rubbing along a post lattice direction, however, did not produce coherent patterning over multiple post spacings (Figure 2c, 2d).  This observation implies a preferred director orientation along a diagonal of the post lattice, with the symmetry of the bistability broken by the weak oriented anchoring potential provided by the rubbing.

\begin{figure}
\includegraphics[width=0.95\linewidth]{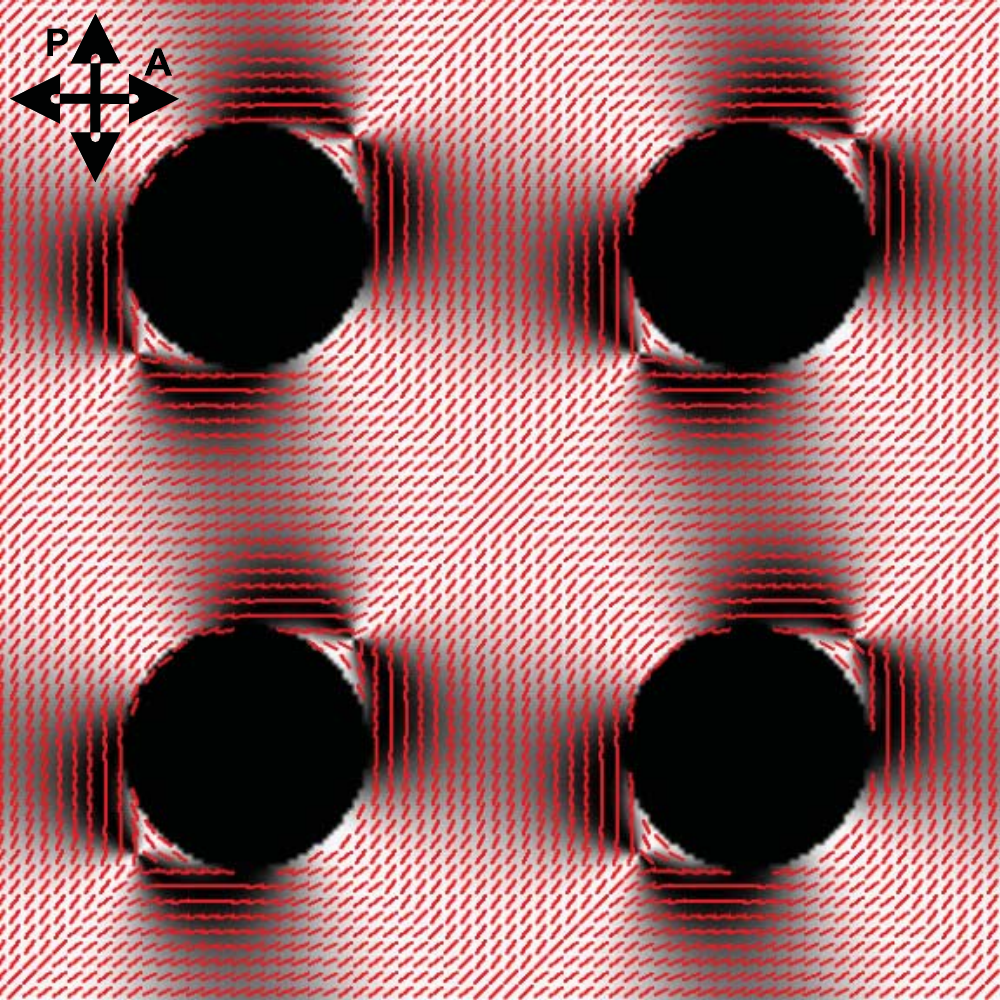}
\centering
\caption{Average in-plane director orientation (red lines) from a minimization of the LdG free energy for a generic nematic in cylindrical micropost confinement, overlaid by the schlieren texture expected when viewed between cross-polarizers.}
\end{figure}

We model a generic nematic film in such a post array using numerical LdG free energy minimization, and we find that the minimum energy configuration does indeed demonstrate average alignment along a diagonal of the micropost lattice, skewing slightly off-diagonal between adjacent posts (Figure 3).  Two vertically aligned -1/2 topological disclinations per post are also observed, separated from the micropost centers along a diagonal of the square lattice, which is parallel to the average director alignment.  This alignment configuration implies a bistable preferred director orientation along either diagonal of the micropost lattice.  We did not observe this configuration in our initial experiments because of the random nucleation of diagonally aligned domains that arises during the isotropic-nematic cooling of the film and thereby creates an ``overall disordered'' director pattern.  The anchoring strength of this rubbing technique, though not strong enough to induce an arbitrary overall alignment, for example, along a lattice  direction (Figure 2c, 2d), is sufficient to break the two-diagonal degeneracy of the system (Figure 2a, 2b).

Quantification of the LC director orientations in the aligned nematic films demonstrates that both DSCG and SSY nematic films have director configurations consistent with our expectations based on the LdG free energy minimizations for a generic nematic LC (Figure 4).  Additionally, the degree of in-plane director order (Figure 4b) is observed to decrease close to the microposts, suggesting the presence of defects or out-of-plane configurations at locations where one would expect -1/2 disclinations in the LdG free energy minimizations.

\begin{figure}
\includegraphics[width=0.95\linewidth]{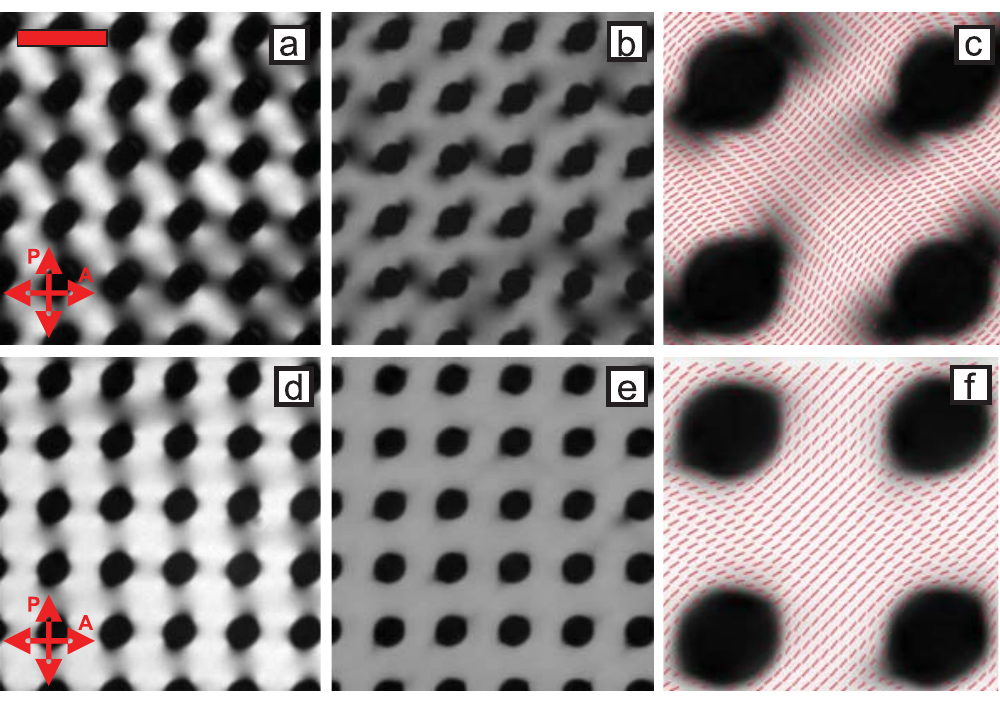}
\caption{Nematic films of 14\% wt/wt DSCG (a-c) and 30\% wt/wt SSY (d-f) in micropost arrays. Images show schlieren textures under cross-polarizers (a \& d), average in-plane director magnitude (b \& e), and average in-plane director orientation for a selected region of 4 posts (c \& f). Scale bar = 20 $\mu$m.}
\end{figure}

\subsection{Columnar Configurations}
Slow drying of these films, the result of water being lost to the overlying hexadecane, transforms the LCLC nematic film into a high-concentration columnar mesophase.  Interestingly, this phase transition is accompanied not only by significant changes in the local director configuration (Figures 5a, 6a, 7a), but also by the appearence of dark lines in bright-field microscope images (Figures 5b, 6b, 7b).  The resulting columnar liquid crystal film forms two very different director configurations dependent on type (i.e., DSCG or SSY) and concentration of the chromonic liquid crystal used to make the initial nematic film.

\begin{figure}
\includegraphics[width=0.95\linewidth]{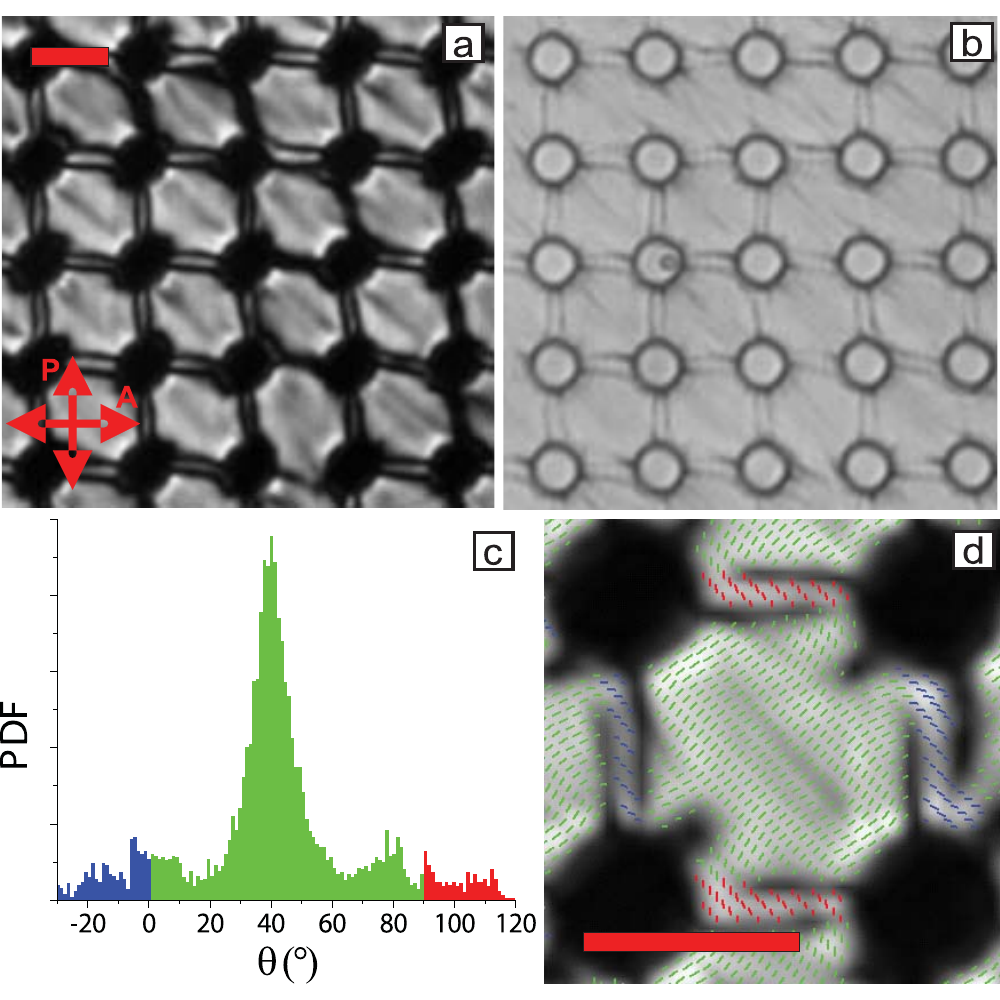}
\caption{Columnar film of DSCG dried from a nematic DSCG film of initial concentration 17\% wt/wt.  Images show (a) cross-polarizer image of the film; (b) bright field image of the film (where defects are visible as dark lines); (c) probability distribution of in-plane director orientations with respect to the horizontal post lattice spacing direction; (d) average in-plane director orientation for four posts, colored by angle using the same scheme as (c) and plotted over the average in-plane director magnitude (bright=in-plane, dark=out-of-plane/disordered). Scale bar = 10 $\mu$m.}
\end{figure}

For columnar DSCG films which cross over from an initial nematic phase with concentration $<$ 17.5\% wt/wt, we observe configurations containing pairs of defect lines running between adjacent posts, forming local squares or rhombuses with posts at the vertices (Figure 5).  By quantifying the director orientation and magnitude (see supplementary information for details), we see that the defect lines in the bright field images correspond to areas with low in-plane director magnitude (Figure 5d).  Additionally, these lines appear to divide the space into regions with distinct (i.e., locally uniform) director orientations.  This observation suggests that the lines are defect walls, i.e., plane-like defects in columnar liquid crystals that occur in regions across which there exists a discontinuous bend in director orientation.\cite{Oswald1982} In these samples, the area between four posts is populated by a director oriented close to a diagonal of a lattice, and the regions between adjacent posts contain directors oriented close to a lattice direction.  Notice that these patterns of alignment are similar to the director configuration found in the nematic films, albeit with a much sharper off-diagonal director bend between adjacent posts.

\begin{figure}
\includegraphics[width=0.95\linewidth]{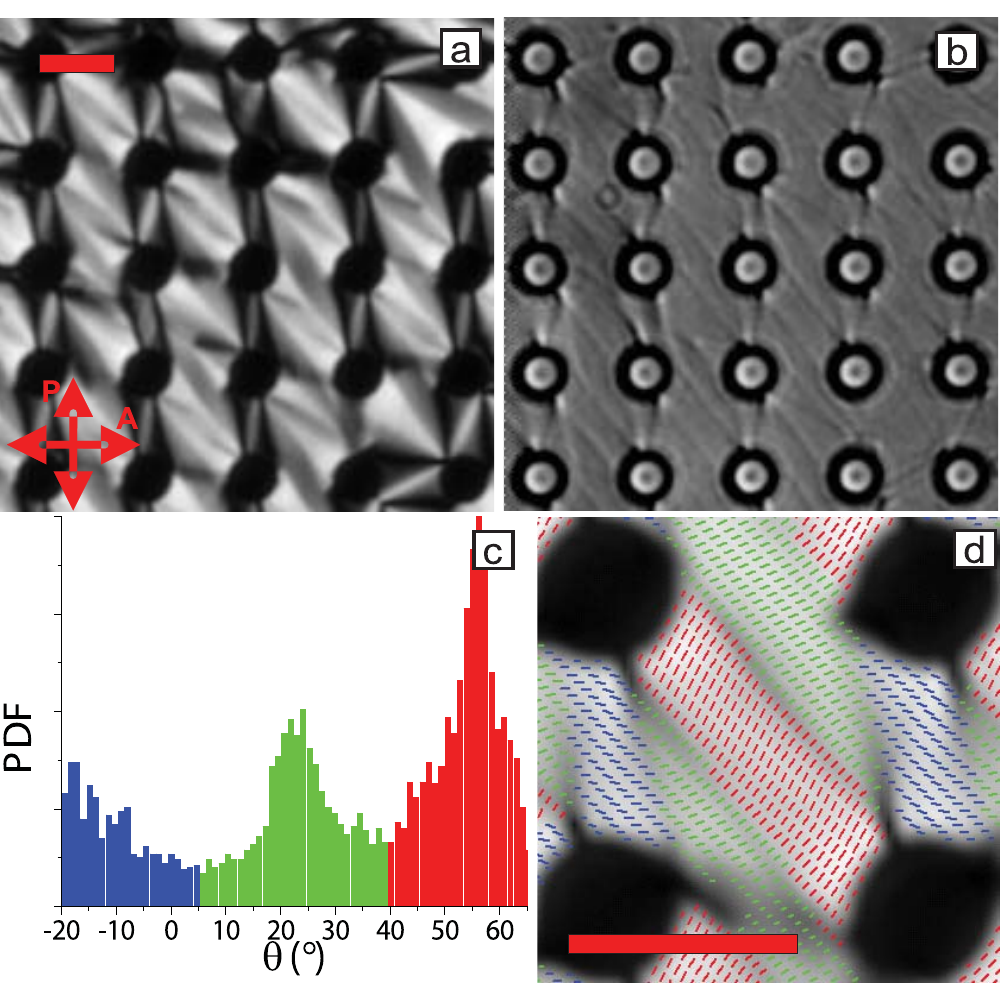}
\caption{Columnar film of DSCG dried from a nematic DSCG film of initial concentration 18\% wt/wt.  Images show (a) cross-polarizer image of the film; (b) bright field image of the film (where defects are visible as dark lines); (c) probability distribution of in-plane director orientations with respect to the horizontal post lattice spacing direction; (d) average in-plane director orientation for four posts, colored by angle using the same scheme as (c) and plotted over the average in-plane director magnitude (bright=in-plane, dark=out-of-plane/disordered). Scale bar = 10 $\mu$m.}
\end{figure}

A different defect placement and local director distribution arises in columnar DSCG films with an initial pre-drying concentration $>$ 18\% wt/wt (Figure 6), and in all columnar SSY films (Figure 7).  The defect walls indicated by dark lines in bright-field (Figures 6b, 7b) and low in-plane director magnitude (Figures 6d, 7d), run between adjacent posts in a single lattice direction, as well as between posts on a lattice diagonal perpendicular to the initial nematic director alignment, forming zig-zags of defect walls.  These defect walls divide the director configurations into three distinct regions.  Between diagonal posts, the director is aligned within 5$^\circ$ of a lattice diagonal; between adjacent posts, the director lies on (or slightly tilted past) a lattice direction; and the remaining rhombus-like region contains a director oriented between these directions, typically 20-30 $^\circ$ off the lattice direction.  Unlike the ``square'' configuration, the average director field in this case appears to lie significantly off the diagonal of the micropost lattice.

\begin{figure}
\includegraphics[width=0.95\linewidth]{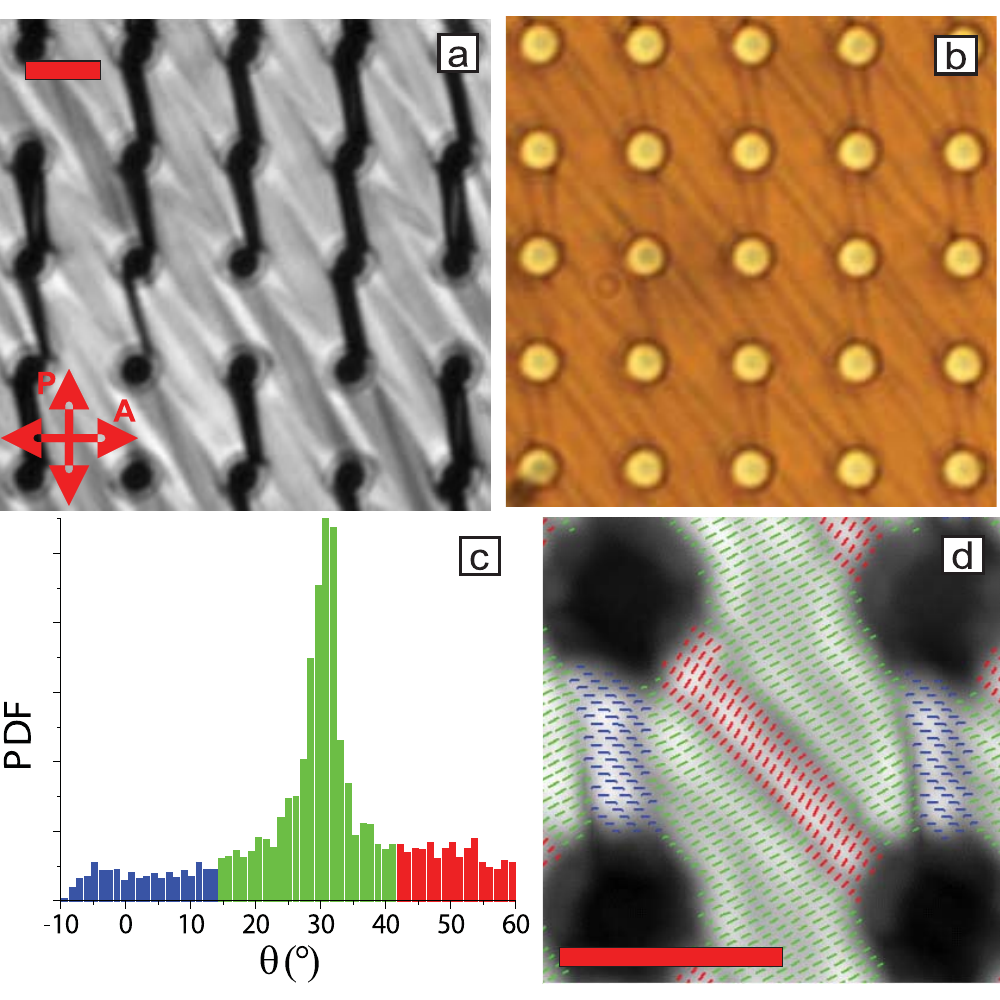}
\caption{Columnar film of SSY.  Images show (a) cross-polarizer image of the film; (b) bright field image of the film (where defects are visible as dark lines); (c) probability distribution of in-plane director orientations with respect to the horizontal post lattice spacing direction; (d) average in-plane director orientation for four posts, colored by angle using the same scheme as (c) and plotted over the average in-plane director magnitude (bright=in-plane, dark=out-of-plane/disordered). Scale bar = 10 $\mu$m.}
\end{figure}

\begin{figure}
\includegraphics[width=0.95\linewidth]{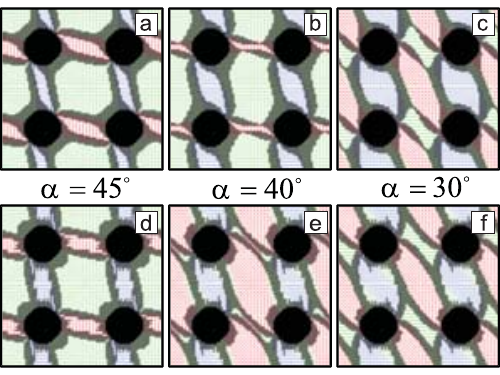}
\caption{Average in-plane director orientations from expanded LdG free energy minimizations in micropillar confinement for effective Frank elastic constant ratios $K_3$/$K_1$ = 0.1 (a-c) and $K_3$/$K_1$ = 0.01 (d-f) for in-plane director initializations of 45$^\circ$ (a \& d), 40$^\circ$ (b \& e), and 30$^\circ$ (c \& f) from the horizontal post lattice spacing direction.  Line coloring highlights distinct regimes of director orientation (blue = close to horizontal, red = most off-horizontal, green = intermediate orientation); darker backgrounds indicate regions with the 40\% greatest contribution to Frank free energy bend, $b=((\textbf{n} \cdot \nabla)\textbf{n})^2$.}
\end{figure}

Numerical modeling of the columnar phase is performed in the low $K_3$/$K_1$ limit of the LdG system described previously. The results indicate that the tendency of different LCLCs at different initial concentrations to form different local patterns is tied to the elastic properties of the LCLCs during the drying process.  The free energy minimizations qualitatively reproduce the director fields exhibited by the columnar LCLC films (Figure 8).  Interestingly, the initialization angle of the director field with respect to the post lattice determines whether the resultant director configuration matches the ``square'' or ``zigzag'' configurations seen in experiment. If the initialization angle for the free energy minimization lies on the lattice diagonal, then a director configuration very similar to the ``square'' configuration is seen; additionally, though no obviously disordered ``defect''-like regions are seen, the areas with highest local bend $b= ((\textbf{n} \cdot \nabla)\textbf{n})^2$, i.e., where a defect wall would most likely occur in a columnar liquid crystal, match qualitatively with the placement of the defect walls seen in experiment.  If the director angle is initialized at 30 $^\circ$ from the post lattice spacing direction (15 $^\circ$ degrees off-diagonal), then a director configuration similar to the ``zigzag'' region is reproduced, with defect walls again predicted by high-bend regions.

We note that variation of the ratio of splay and bend elastic constants $K_1$ and $K_3$, respectively, can change the relative stability of these configurations.  As can be seen in Figure 8, an initialization angle of $\alpha=40^\circ$ with a $K_3$/$K_1$ ratio of 0.1 results in a ``square'' configuration, while the same initialization angle evolving under a $K_3$/$K_1$ ratio of 0.01 produces a ``zigzag'' configuration.  The preferred configuration is described in Figure 9 as a function of the deviation of the initialization angle with respect to the lattice diagonal.  For $K_3$/$K_1$ = 0.1, the ``square'' configuration is stable up to an initialization angle deviation of up to 10$^\circ$ off-diagonal; for $K_3$/$K_1$ = 0.01, this configuration is stable up to a deviation of only 2.5$^\circ$.  This finding implies a decreasing stability of the ``square'' phase and a stability preference of the ``zigzag'' phase as the $K_3/K_1$ ratio decreases.

\begin{figure}
\includegraphics[width=0.95\linewidth]{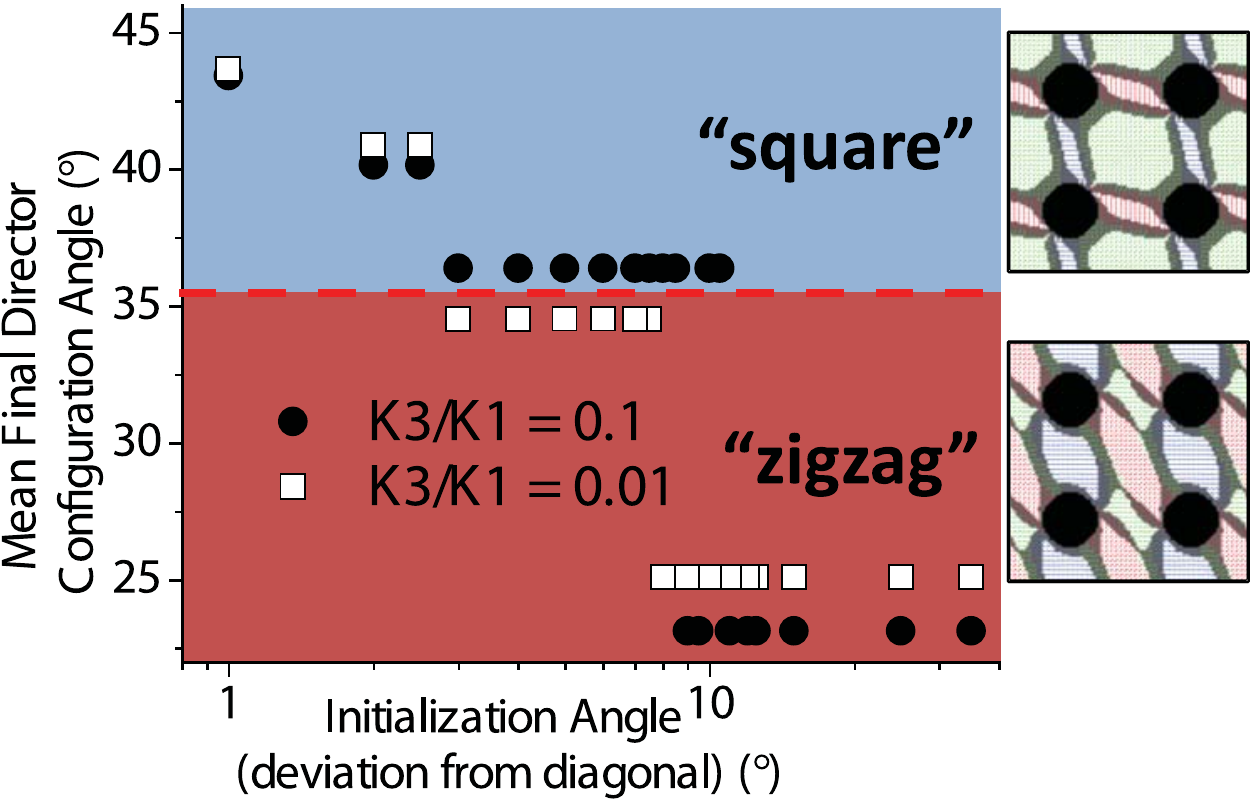}
\caption{Final mean director orientation from expanded LdG free energy minimizations as a function of deviation of initialized director angle from the lattice diagonal.  Mean director angles $>$ 36$^\circ$ correspond to ``square'' defect wall configurations (blue area of graph); mean director angles $<$ 36$^\circ$ correspond to ``zigzag'' defect wall configurations (mauve area of graph).  Filled circles represent effective Frank free energy constants $K_3$/$K_1$ = 0.1; open squares represent effective Frank free energy constants $K_3$/$K_1$ = 0.01}
\end{figure}

This interpretation is consistent with experimental observation and the known elastic properties of the studied LCLCs.  The $K_3$/$K_1$ ratio of nematic DSCG at 14 \% wt/wt is 2.5 at 25 $^\circ$C\cite{Nastishin2008}; for nematic SSY at 29\% wt/wt, the $K_3$/$K_1$ ratio is 1.4, and at 31.5\%, $K_3$/$K_1$ = 0.95.\cite{Zhou2012}   This interpretation is consistent with experimental observation and the known elastic properties of the studied LCLCs.  The $K_3$/$K_1$ ratio of nematic DSCG at 14 \% wt/wt is 2.5 at 25 $^\circ$C\cite{Nastishin2008}; for nematic SSY at 29\% wt/wt, the $K_3$/$K_1$ ratio is 1.4, and at 31.5\%, $K_3$/$K_1$ = 0.95.\cite{Zhou2012}   Though no experimental determinations of DSCG elastic constants at higher nematic concentrations exist, we expect from general expectations about the nematic-columnar transition \cite{Swift1982} and the trends seen in SSY \cite{Zhou2012} that higher concentrations of DSCG should have a lower $K_3$/$K_1$ ratio.  In our experiments, systems which have a higher $K_3$/$K_1$ in the initial nematic configuration (i.e., nematic DSCG at a concentration $<$ 17\% wt/wt) form the ``square'' columnar state.  Likewise, systems which have a lower $K_3$/$K_1$ in the initial nematic configuration (i.e., nematic DSCG at a higher concentration, and nematic SSY) form the ``zigzag'' columnar state.  Though some of the details of the route from nematic to columnar structure formation are not worked out, both the experiments and the numerics suggest the same basic control of columnar configurations via modifications of the elastic constants.

\section{Conclusion}

We have demonstrated control of emergent patterning of a micro-templated aqueous LCLC film by adjusting its elastic properties via variation of the mesogen concentration.  Micropost confinement biases alignment on a preferred direction for nematic films, but columnar films adopt several distinct configurations of defect and director patterns depending on their preparation.  These results demonstrate a novel method for inducing alignment of typically difficult-to-align LCLC films, as well as the first instance of two-dimensional patterning of a columnar liquid crystal film.  These results additionally serve as a prime example of configuration switching in a LC film due to changes in bulk elastic properties, rather than external boundary conditions or applied fields.  Eventually, this work could lead to new ideas about the control of self-assembled patterned films. For example, these particular columnar LCLC films could serve as templates for self-folding materials, since their director ``tiling'' resembles the patterning studied in models of such films.\cite{Modes2011, Modes2013}  Additionally, the patterns and order induced in theses LCLC films could be transferred to suspended colloids, suspended nanotubes and bacterial systems to create novel metamaterial films.\cite{Evans2011, OuldMoussa2013, Smalyukh2008, Mushenheim2014, Zhou2014}

\section{Acknowledgements}

This work is supported by the National Science Foundation through NSF Grant DMR12-05463 and PENN MRSEC Grant DMR11-20901, as well as by NASA through NASA Grant NNX08AO0G.  DAB was supported through NSF Graduate Research Fellowship Grant No. DGE-1321851. This work was partially supported by a Simons Investigator award from the Simons Foundation to RDK. RDK was also supported in part by the Perimeter Institute for Theoretical Physics. We also thank a wide range of colleagues, including Laura Laderman, Zoey Davidson, Joonwoo Jeong, Mohamed Gharbi, Tim Still, Kevin Aptowicz, Ye Xu, and Matthew Gratale, for helpful discussions and technical insight about this and related projects.

\providecommand*{\mcitethebibliography}{\thebibliography}
\csname @ifundefined\endcsname{endmcitethebibliography}
{\let\endmcitethebibliography\endthebibliography}{}

\end{document}